\documentclass[a4paper,11pt]{article}
\pdfoutput=1 

\usepackage{jheppub} 

\usepackage[T1]{fontenc} 

\title{\boldmath Kinetically stabilized inflation}

\author[a]{Changhong Li\footnote{Corresponding Author}}

\affiliation[a]{Department of Astronomy, Key Laboratory of Astroparticle Physics of Yunnan Province,  \\ School of Physics and Astronomy,  Yunnan University, \\ No.2 Cuihu North Road, Kunming,  650091 China}
\emailAdd{changhongli@ynu.edu.cn}
\author[a]{,~ Hao Gong}

\author[b,c]{~and Yeuk-kwan Edna Cheung}

\affiliation[b]{Department of Physics, Physics College,  Nanjing University, \\ No.22 Hankou Road Nanjing,  650091 China}
\affiliation[c]{Institute of Nuclear and Particle Physics,  
Demokritos National Research Centre,   Athens,  Greece.}

\emailAdd{cheung.edna@gmail.com}

\abstract{In this work, we propose a string-inspired two fields inflation model to address the fine-tuning problem that the standard inflation model suffers. The fast-rolling tachyon $\mathcal{T}$ originated from the D-brane and anti-D-brane pair annihilation locks the inflaton $\varphi$ slowly rolling on a Higgs-like potential $V(\varphi)=-m_\varphi^2\varphi^2+\lambda \varphi^4$ and drives a kinetically stabilized (KS) inflation. Our numerical simulation confirms such a solution is a dynamic attractor. In particular, for $\lambda< 0.8\times 10^{-3}$, the e-folding number contributed by the KS inflation phase can be larger than $62$ to solve the horizon and flatness problems of Big Bang theory. Notably, this KS inflation generates a nearly scale-invariant primordial curvature perturbations spectrum consistent with current cosmic microwave background (CMB) observations. It predicts a low tensor-to-scalar ratio, which the current primordial gravitational wave background (the B-modes in CMB) searches favor.}

\begin{document}
\maketitle
\flushbottom

\section{Introduction}    \label{sec:intro}  

The canonical single-field slow-roll inflation (standard inflation) is a paradigm for solving the initial problems of the Big Bang Theory, the flatness and horizon problems \cite{Guth:1980zm, Starobinsky:1980te, Sato:1980yn, Linde:1981mu, Albrecht:1982wi}, and also for generating a nearly scale-invariant primordial curvature perturbation spectrum \cite{Mukhanov:1990me}, which is consistent with current cosmic microwave background (CMB) observations \cite{WMAP:2010qai, Planck:2015fie, Planck:2018vyg}. However, the essential part of the standard inflation, the slow-roll condition, is challenging to realize as it is not a dynamic attractor (see Ref.\cite{Baumann:2009ds} for a review and citations to the original reference). Additionally, the current primordial gravitational wave (PGW) (or primordial B-modes in CMB \cite{Kamionkowski:2015yta}) searches imply that the tensor-to-scalar ratio could be smaller than the conventional expectation in the standard inflation \cite{BICEP:2021xfz, Tristram:2021tvh}. 

This work attempts to realize the inflation process as a dynamic attractor by introducing an auxiliary field. In practice, we adopt a string-inspired open string tachyon field $\mathcal{T}$ to couple with the scalar inflaton $\varphi$ in the Dirac-Born-Infeld (DBI) form. After the open string tachyon condenses following a D-brane and anti-D-brane pair annihilating, the tachyon field locks the inflaton on the not-so-flat inflaton potential to provide a locked slow-roll inflation phase. This proposal has two merits: 1) the D brane and anti-D brane pair annihilation is dynamically irreversible in an expanding universe \cite{Sen:1999nx, Shiu:2002xp}. Therefore, the attractor is stable against various potential shapes and cosmic red-shifts; and 2) after the tachyon condenses into cold matter form \cite{Sen:2002nu, Sen:2002in}, tachyon matter gets red-shifted to sub-dominated without ruining the desired inflation phase.  

According to our analysis, this newly proposed model consists of three phases before the cosmic reheating. 
\begin{enumerate}
    \item {\it Phase I: Annihilating D-brane and anti-D-brane pair dominated inflation.} In this phase, an initial intact D-brane and anti-D-brane pair starts annihilating. Their residual tension drives the Universe to inflate\footnote{This phase is an analogy to the well-known D-brane inflation models (for instance, see Refs. \cite{Dvali:1998pa, Maartens:1999hf, Burgess:2001fx, Kachru:2003sx, Maartens:2003tw} and $k$ inflation models (for instance, see Refs. \cite{Armendariz-Picon:1999hyi, Scherrer:2004au, Babichev:2007dw, Li:2012vta, Cai:2016ngx, Odintsov:2019ahz}). The difference between them is the brane inflation needs a prolonged annihilation process to get a large e-folding number. But in our setup, D-brane and anti-D-brane pair annihilation is very swift as the energy scale is very high and serves as an initial condition for the newly proposed kinetically stabilized inflation.}. Ultimately, the D-brane and anti-D-brane pair releases all energy into the tachyon field and condenses into a cold matter form (the condensed tachyon matter).  
    
    \item {\it Phase II: Tachyon matter dominated expansion.} In this phase, the condensed tachyon matter dominates the cosmic background to expand. As a dynamic attractor (confirmed by our numerical simulation), the fast-rolling tachyon locks the inflaton to roll slowly on a Higgs-like potential, $V(\varphi)=-m_\varphi^2\varphi^2+\lambda\varphi^4$ \footnote{Without loss of generality, we adopt such a Higgs-like potential to facilitate our analysis in this work. Note that this choice differs from the well-known Higgs inflation \cite{Bezrukov:2007ep, Barbon:2009ya, Bezrukov:2010jz, Germani:2010gm} , which field rolls from a large value to a small value. In this work, we assume $\varphi$ rolls down from the false vacuum at $\varphi=0$ (a small value) to the true vacuum $\varphi_v=m_\varphi/\sqrt{2\lambda}$ (a large value).}, behaving like a slow-varying cosmological constant. At the end of this expanding phase, tachyon matter is red-shifted to be sub-dominated, and the slow-roll inflation field $\varphi$ becomes dominant. 
    
    \item {\bf  Phase III: Kinetically stabilized (KS) inflation.} In this phase, the fast-rolling tachyon continues to lock the inflaton to roll down slowly, and the inflation field dominates the evolution of the Universe. We call this inflation phase kinetically stabilized (KS) inflation to distinguish it from standard slow-roll inflation. After a long slow-roll process, the KS inflation eventually attains the true vacuum at $\varphi_v=\sqrt{2}m_\varphi/2\lambda$, and the Universe reheats into standard model (SM) particles dominated era\footnote{In some Higgs warm inflation like Ref. \cite{Dymnikova:2001jy}, the reheating process can happen earlier. In some kinetic axion model $F(R)$ gravity theory like Ref. \cite{Oikonomou:2022tux}, the axion field can oscillate around its vacuum during inflation rather than reheating.} \cite{Bassett:2005xm, Allahverdi:2010xz, Amin:2014eta}. As we will show, taking a generic initial condition for $\varphi$, $\varphi_i\sim m_\varphi$, the e-folding number ($N_e\equiv \ln(a/a_i)$) contributed by the KS inflation (with $\lambda<0.8\times 10^{-3}$  and $m_s=10^{-3}M_p$) can be greater than $62$ to solve the horizon and flatness problems, where $m_s$ being the string mass and $M_p$ the reduced Planck mass. Moreover, by fine-tuning the initial value of $\varphi$ to be $\varphi_i\simeq 10^{-3}m_\varphi$, one can even obtain $\lambda=\mathcal{O}(1)$. We expect many effective field theories to satisfy the requirement of $\lambda$, including the Higgs boson in the SM of particle physics (for a short review, see the Higgs boson section in \cite{ParticleDataGroup:2018ovx}). 
\end{enumerate}

After elucidating the evolution of the cosmic background, we compute the primordial curvature perturbation for the KS inflation. Its spectrum is nearly scale-invariant and consistent with the current observation of E-modes in CMB, similar to the standard inflation, except having a larger pre-factor $c_s^{-1} \gg 1 $. Such a large factor implies a more significant gap between the scalar and tensor modes of primordial metric perturbation for KS inflation. We then compute the tensor-to-scalar ratio for KS inflation and find that compared to the standard inflation, the current PGW searches' results favor KS inflation.   
 
\section{The kinetically coupled tachyon and inflation fields model}
In string theory, the D-brane and anti-D-brane pair annihilation (or the unstable D-branes decay) corresponds to a phase transition from the false vacuum (the open string vacuum) to the true vacuum (the closed string vacuum) \cite{Sen:1999nx}. In this tachyonic process, the D brane and anti-D brane pair releases all energy into the tachyon field $\mathcal{T}$ and make it fully condense into a cold matter form (called tachyon matter) \cite{Sen:2002nu, Sen:2002in}. In the effective theory, one can use the following DBI action of the single tachyon field $\mathcal{T}$ to describe a spatially filling $D_3$ brane and anti-$D_3$ brane pair annihilation \cite{Sen:2004nf},  
\begin{equation}
\mathcal{L}_{\mathcal{T}}=-V(\mathcal{T})\sqrt{1+\partial^\mu \mathcal{T}\partial_\mu \mathcal{T}}~,
\end{equation} 
where the hill-like potential of the tachyon field taking $V(\mathcal{T})=V_0/\cosh(\mathcal{T}/\sqrt{2})$, $V_0=m_s^3M_p$ being the tension of D brane and anti-D brane pair, and $m_s$ and $M_p$ being the string mass and the reduced Planck mass respectively. Throughout this paper, we take $m_s=10^{-3} M_p$ and the convention $ m_s=1$.

In the homogeneous and isotropic cosmic background, which the Friedmann - Lemaitre - Robertson - Walker (FLRW) metric ($g_{\mu\nu}=\{-1, a^2(t)\delta_{ij}\}$) describes, the equation of motion of the tachyon field takes 
\begin{equation}\label{eq:steom}
  \ddot{\mathcal{T}}-3H\dot{\mathcal{T}}w_{st}-\frac{d\ln V(\mathcal{T})}{d\mathcal{T}}w_{st}=0~,
\end{equation}
and the Friedmann equation takes
\begin{equation} \label{eq:fest}
    H^2=\frac{8\pi G}{3}\rho_{st},
\end{equation}
where $w_{st}=-(1-\dot{\mathcal{T}}^2)$ and $\rho_{st}=V(\mathcal{T})/\sqrt{1-\dot{\mathcal{T}}^2}$ being the equation of state (EoS) and the energy density of the single tachyon field respectively. 

Solving Eq.(\ref{eq:steom}) and Eq.(\ref{eq:fest}), one can obtain the cosmic evolution of the tachyon field. In particular, with a small initial displacement, the tachyon field accelerates to roll down from the top of the hill-like potential. This corresponds to the initial intact $D_3$ brane and anti-$D_3$ brane pair ($\rho_\mathcal{T}\simeq V_0$ at $\mathcal{T}\simeq 0$ and $\dot{\mathcal{T}}\simeq 0$ ) starts to annihilate. At the end ($\mathcal{T}\gg 1$), tachyon field attains the maximal velocity 
\begin{equation}\label{eq:stattr}
    \dot{\mathcal{T}}\rightarrow 1 , \ddot{\mathcal{T}}\rightarrow 0~,
\end{equation}
which reflects that the D brane and anti-D brane pair releases all energy into the tachyon field and make it condense into a cold dark matter form \cite{Sen:2004nf}, 
\begin{equation} \label{eq:wstzero}
   w_{st}=-(1-\dot{\mathcal{T}}^2)\rightarrow 0, \quad  \rho_{st}\propto a^{-3},
\end{equation}
where $\rho_{st}$ is the energy density of the single tachyon field.

\begin{figure}[tbph!]
\centering 
\includegraphics[width=0.70\textwidth]{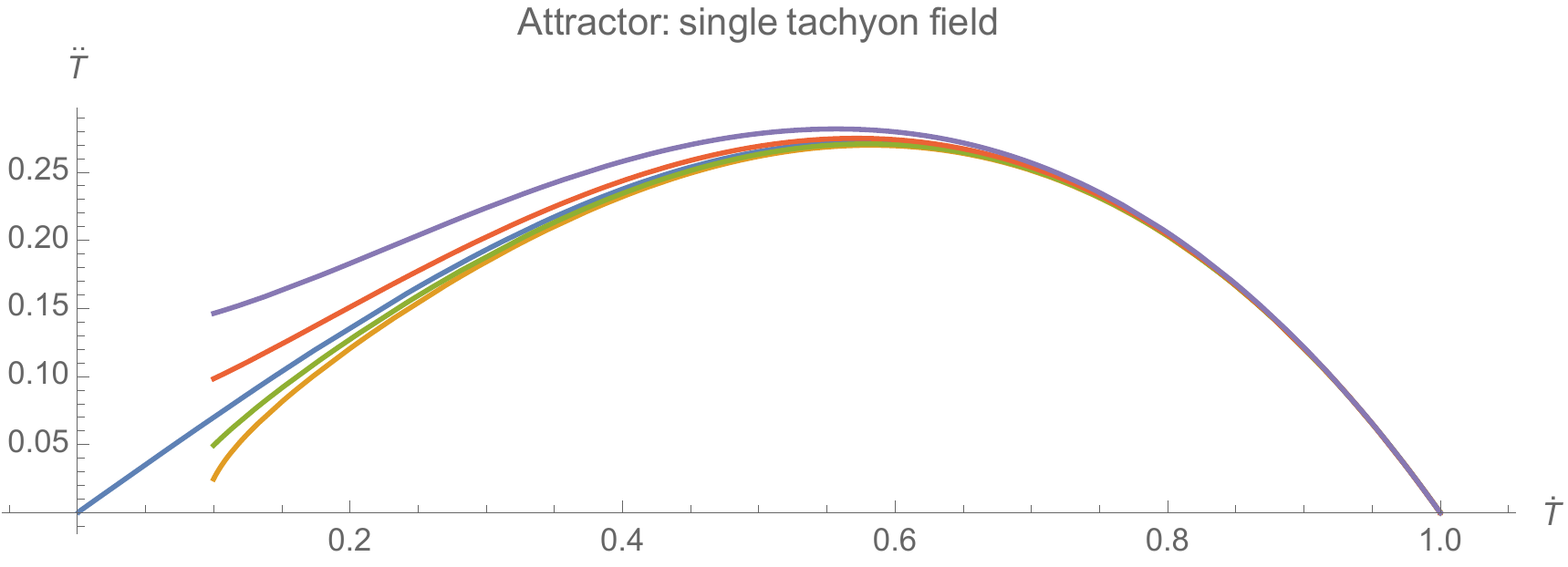}
\caption{\label{fig:Attractor} The parametric plot of $\dot{T}$ and $\ddot{T}$ illustrates that  $(\dot{\mathcal{T}},\ddot{\mathcal{T}})=(1,0)$ is a dynamic attractor for the single tachyon field model. The longest curve is plotted by solving Eq.(\ref{eq:steom}) numerically with initial condition $(\mathcal{T}(0),\dot{\mathcal{T}}(0))=(0.001,0.001) $ while other curves are plotted with $(0.05,0.1)$, $(0.1,0.1)$, $(0.2,0.1)$, and $(0.3,0.1)$ respectively. All curves converge into the fixed point $(\dot{\mathcal{T}},\ddot{\mathcal{T}})=(1,0)$. For simplicity, we have ignored the red-shifted term, $3H$, in this numerical simulation. }
\end{figure}

Encouragingly, such a tachyon condensation process is a dynamic attractor. Figure \ref{fig:Attractor} illustrates tachyon condensation in the phase space. $(\dot{\mathcal{T}},\ddot{\mathcal{T}})=(1,0)$ is a dynamic attractor for the tachyon field evolution, which reflects D-brane and anti-D-brane pair annihilating into tachyon matter is inevitable and irreversible. In the D-brane/k inflation and the matter bounce cosmology (for instance, see Refs. \cite{Dvali:1998pa, Maartens:1999hf, Burgess:2001fx, Kachru:2003sx, Maartens:2003tw, Armendariz-Picon:1999hyi, Scherrer:2004au, Babichev:2007dw, Li:2012vta, Cai:2016ngx, Odintsov:2019ahz} and Refs. \cite{Sen:2003mv, Li:2011nj, Li:2013bha, Zhang:2019tct} respectively), such a single tachyon condensation process has been extensively studied.  

In this work, we adopt this dynamic attractor ($\dot{\mathcal{T}}\rightarrow 1$, $\ddot{\mathcal{T}}\rightarrow 0$) to lock the inflaton $\varphi$ to slowly roll ($\dot{\varphi}\simeq 0$)  on a not-so-flat potential and avoid the fine-tuning problem of the standard inflation. In practice, we consider the following simple extension,
\begin{equation}
\mathcal{L}_{\text{kinetic}}=-V(\mathcal{T})\sqrt{1+\partial^\mu \mathcal{T}\partial_\mu \mathcal{T}+\partial^\mu \varphi\partial_\mu \varphi}~.
\end{equation} 
Without loss of generality, the not-so-flat potential of $\varphi$ takes a Higgs-like form,
\begin{equation}
V(\varphi)=-m_\varphi^2\varphi^2+\lambda \varphi^4~,
\end{equation}
where $m_\varphi$ being the mass of $\varphi$, and $\lambda$ being a dimensionless coupling constant. In this work, we assume that $m_\varphi\ll m_s\ll M_p$.

 In this work, we consider that the Universe is dominated by these coupled tachyon and scalar fields ($\mathcal{L}_{\text{tot}}=\mathcal{L}_{\text{kinetic}}-V(\varphi)$). The tachyon field part describes a spatially filling $D_3$ brane and anti-$D_3$ brane pair, which will annihilate soon and condense into a form of massive cold tachyon matter. The scalar field part, which could be identified with inflaton in this model, is assumed to couple with the tachyon field in the DBI form by its kinetic term. Initially, the scalar field can be viewed as an extra external field weakly acting on the $D_3$ brane and anti-$D_3$ brane pair. After the $D_3$ brane and anti-$D_3$ brane pair annihilating, the massive tachyon matter locks it to realize a slow-roll inflation \footnote{This model can be viewed as one of the specific realizations in the generalized multi-fields Horndeski model \cite{Horndeski:1974wa, Nicolis:2008in, Deffayet:2009wt, Deffayet:2009mn, Deffayet:2011gz, Kobayashi:2011nu, Charmousis:2011bf, Gleyzes:2014dya, BenAchour:2016fzp, Crisostomi:2016tcp, deRham:2016wji, Kobayashi:2019hrl}. We thank the anonymous referee for pointing this out to us. }. Varying the action of this newly proposed model, 
\begin{equation}\label{eq:actiontot}
S=\int dx^{4}\sqrt{-g}\left[\frac{1}{16\pi G}R+\mathcal{L}_{\text{kinetic}}-V(\varphi)\right],
\end{equation}
with respect to  $g^{\mu\nu}$, $\mathcal{T}$ and $\varphi$ respectively, we obtain 
\begin{equation}\label{eq:efeforcth}
R_{\mu\nu}-\frac{1}{2}g_{\mu\nu}R=8\pi G \left[
\left(\partial_\mu \mathcal{T}\partial_\nu \mathcal{T}+\partial_\mu \varphi\partial_\nu \varphi\right)\rho_\mathcal{T}+g_{\mu\nu}\left(p_\mathcal{T}-V(\varphi)\right)\right],
\end{equation} 
\begin{equation}\label{eq:eofmt}
\partial_\mu\left[\sqrt{-g}\rho_\mathcal{T}\partial^\mu \mathcal{T}\right]+\sqrt{-g}p_\mathcal{T}\frac{d\ln V(\mathcal{T})}{d\mathcal{T}}=0,
\end{equation}
and
\begin{equation}\label{eq:eofmvp}
\partial_\mu\left[\sqrt{-g}\rho_\mathcal{T}\partial^\mu \varphi\right]-\sqrt{-g}\left(\frac{dV(\varphi)}{d\varphi}\right)=0,
\end{equation}
where $\rho_\mathcal{T}\equiv V(\mathcal{T})/\sqrt{1+\partial^\mu \mathcal{T}\partial_\mu \mathcal{T}+\partial^\mu \varphi\partial_\mu \varphi}$ and 
$p_\mathcal{T}\equiv-V(\mathcal{T})\sqrt{1+\partial^\mu \mathcal{T}\partial_\mu \mathcal{T}+\partial^\mu \varphi\partial_\mu \varphi}$
being the energy density and the pressure of the coupled tachyon field.

\section{\label{sec:bgev}Kinetically stabilized inflation}

At the background order, Eq.(\ref{eq:efeforcth}), Eq.(\ref{eq:eofmt}) and Eq.(\ref{eq:eofmvp}), respectively, take
\begin{equation}\label{eq:totH}
    H^2=\frac{8\pi G}{3}\left[\rho_{\mathcal{T}}+V(\varphi)\right]~,
\end{equation}

\begin{equation} \label{eq:eofbacktn}
   \ddot{\mathcal{T}}-3H\dot{\mathcal{T}}\frac{w_\mathcal{T}}{1-\dot{\varphi}^2}-\frac{d \ln[V(\mathcal{T})]}{dt}\dot{\mathcal{T}}^{-1}w_\mathcal{T}- \dot{\varphi}\ddot{\varphi}\dot{\mathcal{T}}\frac{1}{1-\dot{\varphi}^2} =0~,
\end{equation}
and \footnote{Note that, for the coupled tachyon field, the attractor is at $\dot{\mathcal{T}}^2=1-\dot{\varphi}^2$ in contrast to the single tachyon field at $\dot{\mathcal{T}}^2=1$. Therefore, the last term of Eq.(\ref{eq:eofbackvpn}) is not divergent at the attractor. }

\begin{equation} \label{eq:eofbackvpn}
    \ddot{\varphi}-3H\dot{\varphi}\frac{w_\mathcal{T}}{1-\dot{\mathcal{T}}^2}-\frac{d \ln[V(\mathcal{T})]}{dt}\dot{\varphi}\frac{w_\mathcal{T}}{1-\dot{\mathcal{T}}^2}-\dot{\mathcal{T}}\ddot{\mathcal{T}}\dot{\varphi}\frac{1}{1-\dot{\mathcal{T}}^2}-\frac{w_\mathcal{T}}{1-\dot{\mathcal{T}}^2}\rho_\mathcal{T}^{-1}\frac{d V(\varphi)}{d\varphi}=0~.
\end{equation}
where $w_\mathcal{T}\equiv p_\mathcal{T}/\rho_\mathcal{T}=-(1-\dot{\mathcal{T}}^2-\dot{\mathcal{\varphi}}^2)$ being the EoS of the coupled tachyon field. Afterward, we use $\mathcal{T}$ and $\varphi$ to denote the background field and $\delta \mathcal{T}$ and $\delta\varphi$ to denote their perturbations.  

Solving Eq.(\ref{eq:totH}), Eq.(\ref{eq:eofbacktn}) and Eq.(\ref{eq:eofbackvpn}) with the initial condition ($\mathcal{T}=\mathcal{T}(0)\ll 1$, $\dot{\mathcal{T}}=0$, $\varphi=\varphi(0)\simeq m_\varphi$, $\dot{\varphi}=0$, $a=a(0)$, and $\dot{a}=0$ at $t=0$), we obtain the following three cosmic phases.

\begin{enumerate}
\item {\bf Annihilating D-brane and anti-D-brane pair dominated inflation.} Before fully annihilating, the tension of the D-brane and anti-D-brane pair ($\rho_\mathcal{T}\simeq V_0$ at $t\simeq 0$) dominates the Universe to inflate for a short period, $\Delta \Tilde{t}\simeq \mathcal{O}(1)\times m_s^{-1}$ \footnote{By fine-tuning the initial values of $\mathcal{T}$ to be very tiny, for instance, $\mathcal{T}(0)=10^{-20} m_s$, the annihilating D-brane and anti-D-brane pair dominated inflation phase can last into a much longer period \cite{Li:2011nj}. But such a setup is too artificial. In this work, we consider a generic initial condition $\mathcal{T}(0)\simeq 10^{-3} m_s$.},
\begin{equation}
a(t)\simeq a(0) e^{\sqrt{\frac{8\pi G}{3}V_0}t}~.
\end{equation}
During this process, the coupled tachyon field accelerates to its maximal velocity, 
\begin{equation}\label{eq:mvofct}
    \dot{\mathcal{T}}\rightarrow \sqrt{1-\dot{\varphi}^2}~, \quad \ddot{\mathcal{T}}\rightarrow 0,
\end{equation}
and condensates into cold matter, which is analogy to Eq.(\ref{eq:wstzero}),
\begin{equation}\label{eq:eosctf}
    w_\mathcal{T}=-(1-\dot{\mathcal{T}}^2-\dot{\mathcal{\varphi}}^2)\rightarrow 0~.
\end{equation}
Note that the D-brane and anti-D-brane pair annihilates very swiftly ($\Delta \Tilde{t}\simeq \mathcal{O}(1)\times m_s^{-1}$ as aforementioned), so the e-folding number contributed by this phase is small, $N_e=\ln{a(\Delta \Tilde{t})/a(0)}$ $\simeq \sqrt{m_s/M_p}\ll 1$, and the value of $\varphi$ is almost unchanged (This is because the slop of $\varphi$ potential is much smaller than tachyon field, $m_\varphi^2\ll m_s^2$. ),
\begin{equation}\label{eq:vpdvpda}
\varphi(\Delta \Tilde{t})\simeq \varphi(0)e^{\mathcal{O}(1)\times\frac{m_\varphi^2}{m_s\sqrt{M_p m_s}}}\simeq \varphi(0)\simeq m_\varphi~, \dot{\varphi}(\Delta \Tilde{t})\simeq\frac{m_\varphi^2}{\sqrt{m_s M_p}}\varphi(\Delta \Tilde{t})\ll m_s^2=1
\end{equation}
where $m_\varphi\ll m_s\ll M_p$ being used. Substituting Eq.(\ref{eq:vpdvpda}) into Eq.(\ref{eq:mvofct}), we obtain a dynamic attractor solution, which is an analogy to Eq.(\ref{eq:stattr}),
\begin{equation}\label{eq:dattct}
    \dot{\mathcal{T}}\rightarrow 1~, \quad \ddot{\mathcal{T}}\rightarrow 0.
\end{equation}

In Figure \ref{fig:tattractor} and Figure \ref{fig:phiattractor}, we numerically solve Eq.(\ref{eq:eofbacktn}) and Eq.(\ref{eq:eofbackvpn}) to illustrate this dynamic attractor for $\mathcal{T}$ part and $\varphi$ part respectively. For simplicity, we have neglected the red-shifted term, $3H$, in the expanding phase and neglected the $\lambda \varphi^4$ term near $\varphi\simeq 0$ region\footnote{We take $\varphi(0)\simeq 0.01 m_\varphi$ and $m_\varphi=10^{-6}m_s$ ($\varphi(0)\sim 10^{-8}$) to perform this simulation. One has plenty of computing resources that can reproduce similar numerical results for other initial values.}. Clearly, the coupled tachyon field $\mathcal{T}$ is also bound to move toward the dynamic attractor, $(\dot{\mathcal{T}},\ddot{\mathcal{T}})=(1,0)$, as shown in Figure \ref{fig:tattractor}. And the inflation field $\varphi$ is locked around the slow-roll condition, $\dot{\varphi}=0$, as shown in Figure \ref{fig:phiattractor}. Note that without such a dynamic attractor, $\varphi$ will increase exponentially and destroy the slow-roll condition. After the D-brane and anti-D-brane pair fully annihilates, the Universe evolves into the next phase. 

\begin{figure}[tbph!]
\centering 
\includegraphics[width=0.70\textwidth]{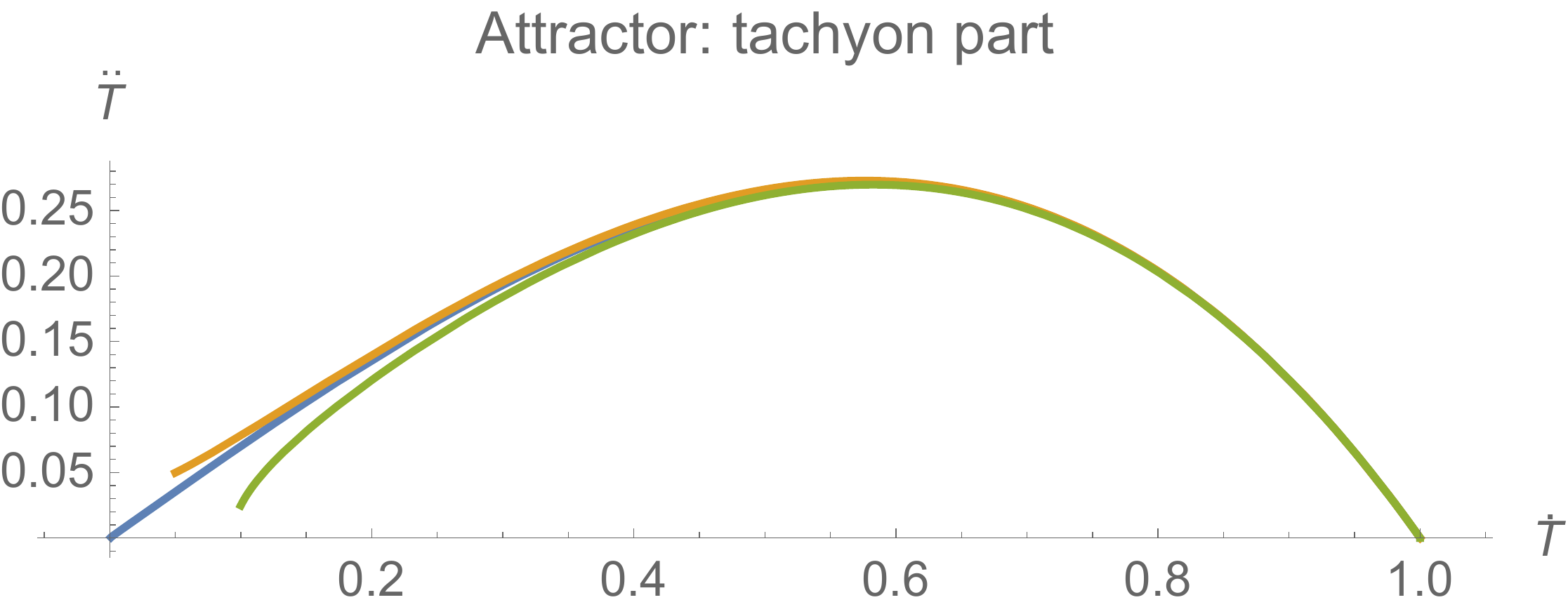}
\caption{\label{fig:tattractor} The parametric plot of $\dot{\mathcal{T}}$ and $\ddot{\mathcal{T}}$ obtained by numerically solving Eq.(\ref{eq:eofbacktn}) and Eq.(\ref{eq:eofbackvpn}), which illustrates that $(\dot{\mathcal{T}},\ddot{\mathcal{T}})=(1,0)$ is a dynamic attractor for the coupled tachyon field. The longest curve is plotted with the initial condition $(\mathcal{T}(0),\dot{\mathcal{T}}(0),\varphi(0),\dot{\varphi}(0))=(0.001,0.001,10^{-8},0) $ and the other two curves are plotted with $(0.1,0.05,3\times 10^{-8},0) $ and $(0.05,0.1,2\times 10^{-8},0) $ respectively. In this numerical simulation, we have neglected the red-shifted term ($3H$) for the expanding phase and neglected the $\lambda \varphi^4$ term near $\varphi\simeq 0$ region for simplicity. }
\end{figure}

\begin{figure}[tbph!]
\centering 
\includegraphics[width=0.70\textwidth]{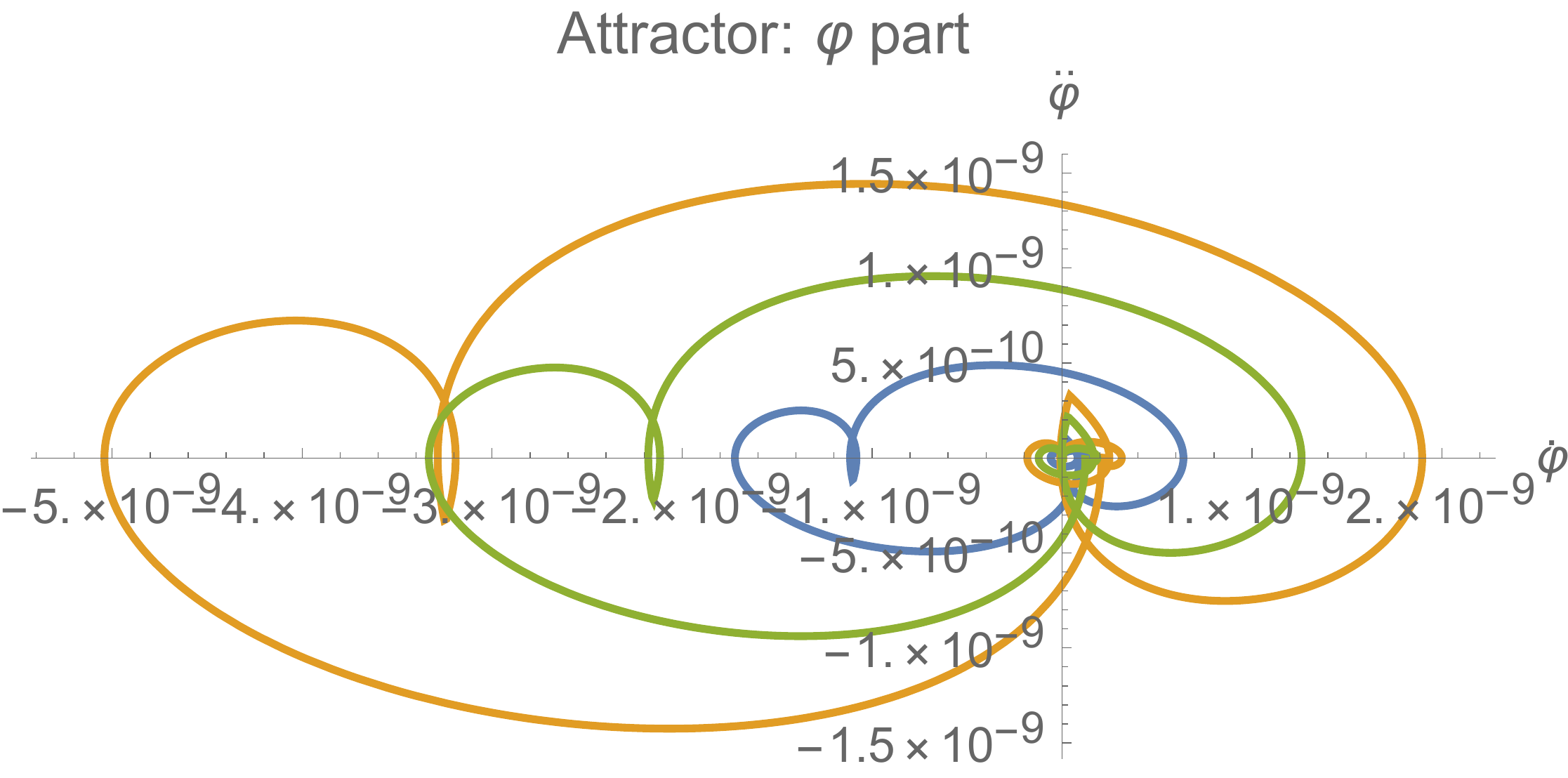}
\caption{\label{fig:phiattractor} The parametric plot of $\dot{\varphi}$ and $\ddot{\varphi}$ obtained by numerically solving Eq.(\ref{eq:eofbacktn}) and Eq.(\ref{eq:eofbackvpn}), which illustrates that the dynamic attractor of the coupled tachyon field ($(\dot{\mathcal{T}},\ddot{\mathcal{T}})=(1,0)$) can lock $\varphi$ around the slow-roll condition $\dot{\varphi}=0$. The smallest curve is plotted with the initial condition $(\mathcal{T}(0),\dot{\mathcal{T}}(0),\varphi(0),\dot{\varphi}(0))=(0.001,0.001,10^{-8},0) $ and the other two curves are plotted with $(0.1,0.05,3\times 10^{-8},0) $ and $(0.05,0.1,2\times 10^{-8},0) $ respectively. In this numerical simulation, we have neglected the red-shifted term ($3H$) for the expanding phase and neglected the $\lambda \varphi^4$ term near $\varphi\simeq 0$ region for simplicity.}
\end{figure}

\item {\bf Tachyon matter dominated expansion.} After the D-brane and anti-D-brane pair annihilation completing,  the condensed coupled tachyon matter dominates the Universe, 
\begin{equation}
    H^2\simeq \frac{8\pi G}{3}\left[V_0\left(\frac{a(\Delta \Tilde{t})}{a(t)}\right)^3+\rho_\varphi\right]\simeq \frac{8\pi G V_0}{3}\left(\frac{a(\Delta \Tilde{t})}{a(t)}\right)^3~,
\end{equation}
where the energy density of $\varphi$ taking \footnote{Taking $V(\varphi_v)=0$ at true vacuum ($\varphi=\varphi_v$), the energy density of $\varphi$ around the false vacuum ($\varphi=0$) takes 
\begin{equation}
    \rho_\varphi=\frac{1}{2}\dot{\varphi_i}^2+\frac{m_\varphi^4}{4\lambda}+\lambda\varphi_i^4 \simeq \frac{m_\varphi^4}{4\lambda},
\end{equation} 
where $\frac{m_\varphi^4}{4\lambda}$ being the $V(\varphi)$ difference between true and false vacuum, and $\lambda<1$ and Eq.(\ref{eq:vpdvpda}) being used.},
\begin{equation}
    \rho_\varphi\simeq \frac{m_\varphi^4}{4\lambda}.
\end{equation}
At the beginning of this phase, $\rho_\varphi$ is sub-dominated, $\rho_\varphi\ll \rho_\mathcal{T}(t)\simeq V_0\left[a(\Delta \Tilde{t})/a(t)\right]^3$. However, the aforementioned dynamic attractor locks $\varphi$ to roll slowly, $\dot{\varphi}\simeq 0$. Thus $\rho_\varphi$ behaves as a slow-varying cosmological constant against the red-shift effect and eventually dominates at the end of this phase, $\rho_\varphi\ge \rho_\mathcal{T}(t_f) \simeq V_0\left[a(\Delta \Tilde{t})/a_f\right]^3$, where the subscript $~_f$ denoting the end of this phase and the beginning of KS inflation.

\item {\bf Kinetically stabilized inflation.} In this phase,  $\rho_\varphi$ is dominant,
\begin{equation}
    H^2=\frac{1}{3M_p^2}\frac{m_\varphi^4}{4\lambda}~,
\end{equation}
which solution takes
\begin{equation}\label{eq:ksiae}
    a=a_f e^{\frac{1}{2\sqrt{3}}\frac{m_\varphi^2}{\sqrt{\lambda}M_p}(t-t_f)}~.
\end{equation}
The dynamic attractor locks $\varphi$ around $\dot{\varphi}\simeq 0$ to realize slow-roll inflation, the kinetically stabilized (KS) inflation. As shown in Figure \ref{fig:tattractor} and Figure \ref{fig:phiattractor}, KS inflation does not suffer the fine-tuning problem of the standard inflation. 

During this phase, $\varphi$ rolls very slowly, so it takes a long time to attain its true vacuum, $\varphi_v=m_\varphi/\sqrt{2\lambda}$. Then the Universe reheats into the SM particles-dominated era. In Appendix \ref{B}, we estimate the e-folding number contributed by the KS inflation phase and obtain
\begin{equation}\label{eq:nefold}
    N_e=\ln\left(\frac{a_r}{a_f}\right)\simeq \frac{1}{4\sqrt{2}}\frac{1}{\lambda}\sqrt{\frac{m_s}{M_p}}\left(\frac{m_\varphi}{\varphi_i}\right)\simeq\frac{1}{4\sqrt{2}}\frac{1}{\lambda}\sqrt{\frac{m_s}{M_p}}~,
\end{equation}
where $a_r$ denotes the end of KS inflation and the onset of cosmic reheating. We adopt a generic initial value for $\varphi$, $\varphi_i\simeq m_\varphi$, rather than fine-tuning it. Note that this generic condition ($\varphi_i\simeq m_\varphi$) coincidentally cancels other $m_\varphi$ factors in Eq.(\ref{eq:nefold}). This fact reflects our analysis neglecting the possibility of $\varphi(0)$ being suspiciously small.   

For solving the horizon and flatness problems, the e-folding number should be larger than $62$ \cite{Dodelson:2003ft}. In Figure \ref{fig:CoPlotNe}, we use Eq.(\ref{eq:nefold}) to plot the parameter region of $(m_s,\lambda)$ that allows $N_e\ge 62$ for the KS inflation. In particular, to ensure $N_e\ge 62$, we obtain $\lambda \le 0.8\times 10^{-3}$ with $m_s/M_p=10^{-3}$. Note that according to the third term in Eq.(\ref{eq:nefold}), we can even obtain $\lambda=\mathcal{O}(1)$ by fine-tuning the initial value of $\varphi$ to be $\varphi_i\simeq 10^{-3}m_\varphi$. We expect many effective field theories can satisfy the requirement of $\lambda$, including the Higgs boson in the SM of particle physics \cite{ParticleDataGroup:2018ovx}. In the next section, we are computing the power spectrum of primordial curvature perturbation generated during this KS inflation phase.  
\begin{figure}[tbph!]
\centering 
\includegraphics[width=0.70\textwidth]{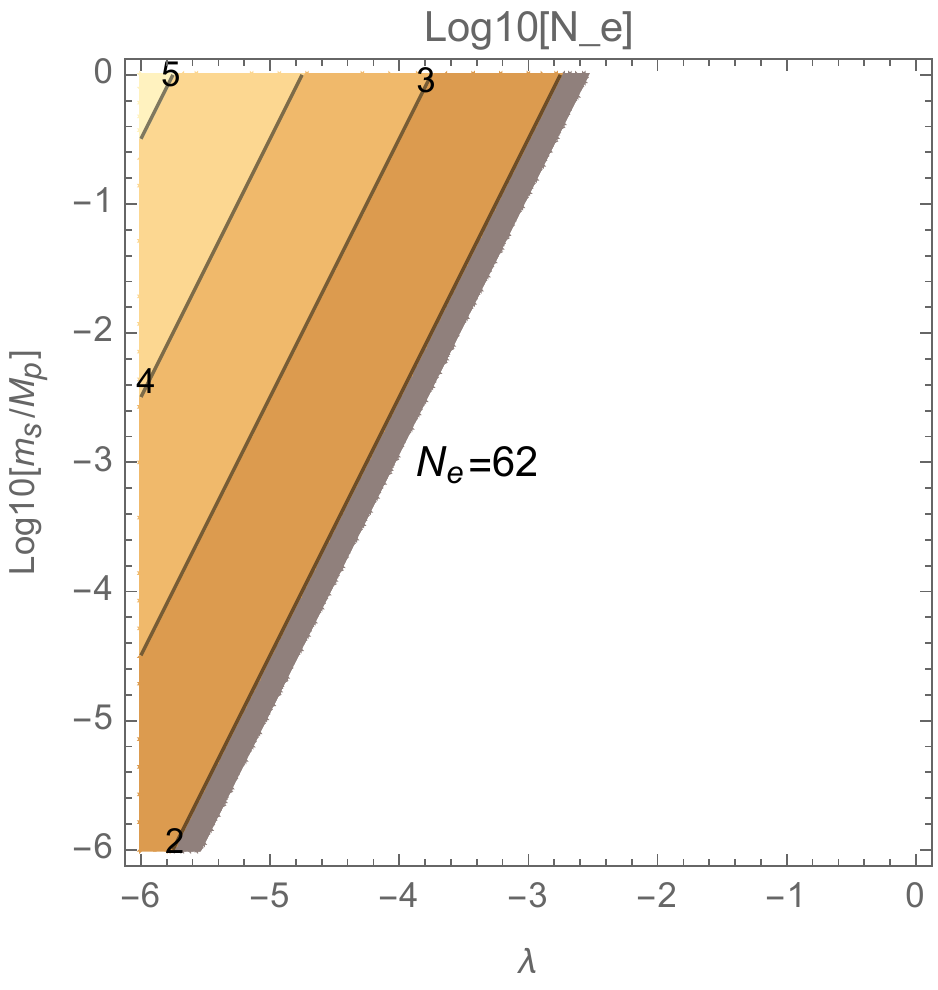}
\caption{\label{fig:CoPlotNe} The contour plot of $N_e$ in $(m_s,\lambda)$-plane for $N_e\ge 62$ in the KS inflation, using Eq.(\ref{eq:nefold}).}
\end{figure}

\end{enumerate}

\section{\label{sec:sipcs}The nearly scale-invariant primordial curvature perturbation spectrum of kinetically stabilized inflation}

As presented in Appendix~\ref{C}, adopting the perturbative Friedmann-Lemaitre-Robertson-Walker (FLRW) metric in conformal Newtonian gauge, 
\begin{equation} \label{eq:flrwall}
g_{\mu\nu}=\{-1-2\Psi(\vec{x},t), a^2(t)\delta_{ij}\left[1+2\Phi(\vec{x},t)\right]\}~,
\end{equation}
and expanding Eq.(\ref{eq:efeforcth}) with $\mathcal{T}=\mathcal{T}+\delta\mathcal{T} $
and $\varphi=\varphi+\delta\varphi$ to the first order, we obtain
\begin{equation}\label{eq:phikeqfull}
\ddot{\Phi}_k+\left(H-\frac{\ddot{H}}{\dot{H}}\right)\dot{\Phi}_k+\left(2\dot{H}-H\frac{\ddot{H}}{\dot{H}}\right)\Phi_k+\frac{c_s^2k^2}{a^2}\Phi_k=0~,
\end{equation}
where $\Phi_k$ being the Fourier modes of $\Phi(\vec{x},t)$ and $c_s^2=1-\dot{\mathcal{T}}^2-\dot{\varphi}^2$ being the sound speed parameter. Note that  Eq.(\ref{eq:phikeqfull}) is the same as the standard inflation ({\it c.f.} Eq.), except $c_s \ll 1$ for the KS inflation. It implies that the KS inflation also can generate a nearly scale-invariant power spectrum.  


Adopting the gauge-invariant variable, $\zeta=\Phi-\frac{H}{\dot{H}}\left(\dot{\Phi}+H\Phi\right)$,
and a canonical variable, $v=z\zeta$ with $z\equiv a\sqrt{2\epsilon}c_s^{-1}$ and $\epsilon\equiv -\dot{H}/H^2$ \cite{Mukhanov:1990me}, we simplifies Eq.(\ref{eq:phikeqfull}) to be 
\begin{equation}\label{eq:vkgeneral}
v_k^{\prime\prime}+\left(c_s^2k^2-\frac{z^{\prime\prime}}{z}\right)v_k=0
\end{equation}
where the superscript $~^\prime$ denoting the derivative with respect to the conformal time $\eta=\int a^{-1}dt$. During KS inflation, $\eta=(aH)^{-1}\propto a^{-1}$ and $z^{\prime\prime}/z=2\eta^{-2}$, which leads to
\begin{equation}\label{eq:vkinflation}
v_k^{\prime\prime}+\left(c_s^2k^2-\frac{2}{\eta^2}\right)v_k=0~.
\end{equation}
Solving Eq.(\ref{eq:vkinflation}) within the Bunch-Davies vacuum ($v_k=\frac{e^{-ic_sk\eta}}{\sqrt{2c_sk}}$ for $\eta\rightarrow -\infty$), we obtain 
\begin{equation}
v_k=\frac{e^{ic_s k\eta}}{\sqrt{2c_s k}}\left(1-\frac{i}{c_s k\eta}\right)~,
\end{equation}
and its spectrum for KS inflation,
\begin{equation}\label{eq:pcpksi}
\mathcal{P}_\zeta\equiv \frac{k^3}{2\pi^2 M_p^2}|\zeta_k|^2=\frac{k^3}{2\pi^2}\left|\frac{v_k}{z}\right|^2=c_s^{-1}\times\frac{H^2}{8\pi^2 M_p^2 |\epsilon|},\quad k\eta\rightarrow 0.
\end{equation}
Eq.(\ref{eq:pcpksi}) indicates that the KS inflation also has a nearly scale-invariant primordial curvature spectrum and is consistent to current observations of the E-modes in CMB, similar with the standard inflation, except $c_s^{-1}\gg 1$. As we will show in the next section, the larger pre-factor $c_s^{-1}$ implies that the KS inflation has a more significant gap between the scalar and the tensor modes spectra, {\it i.e.} a smaller tensor-to-scalar ratio compared to the standard inflation.

\section{The tensor-to-scalar ratio of KS inflation}
During KS inflation (Eq.(\ref{eq:ksiae})), solving the equation of motion for the Fourier tensor modes of metric perturbation $h_{+/\times}$,
\begin{equation}
h^{\prime\prime}_k+2\frac{a^\prime}{a}h^\prime_k+k^2h_k=0~
\end{equation}
we obtain the solution,
\begin{equation}
h_k=\frac{\sqrt{16\pi G}}{a} \frac{e^{-ik\eta}}{\sqrt{2k}}\left(1-\frac{i}{k\eta}\right)~,
\end{equation}
and the spectrum, 
\begin{equation}\label{eq:hspksi}
\mathcal{P}_h\equiv \frac{k^3}{2\pi^2}|h_k|^2=\frac{H^2}{2\pi^2 M_p^2}~, \quad k\eta\rightarrow 0,
\end{equation}
which is identical to the well-known expression in the standard inflation \cite{Dodelson:2003ft}.

Using Eq.(\ref{eq:pcpksi}) and Eq.(\ref{eq:hspksi}), we obtain the tensor-to-scalar ratio for the KS inflation,
\begin{equation}
    r\equiv \frac{\mathcal{P}_h}{\mathcal{P}_\xi}=c_s\times 4|\epsilon|,
\end{equation}
which is the same as the standard inflation, except $c_s\ll 1$. Using the data $\epsilon \simeq(1-n_s)/4 \simeq 0.01$ and $n_s\simeq 0.96$ from WMAP and Planck observations \cite{WMAP:2010qai, Planck:2015fie, Planck:2018vyg}, we find that compared to the standard inflation ($r_\textbf{standard}=4\epsilon\simeq 0.04$), the newly proposed KS inflation ($r_\textbf{KSI}=c_s\times 4|\epsilon| \ll 0.04$) is favored by the current PGW searches ($r_\textbf{BICEP+}< 0.03$) \cite{BICEP:2021xfz, Tristram:2021tvh}. 

\section{Conclusion} 
We proposed the kinetically stabilized (KS) inflation model in this paper. In this model, the fast-rolling tachyon field originated from the D-brane and anti-D-brane pair annihilation locks the inflaton slowly rolling on a Higgs-like potential ($V(\varphi)=-m_\varphi^2\varphi^2+\lambda \varphi^4$) and drives a long inflation process ($N_e>62$ for $\lambda<0.8\times 10^{-3}$). Moreover, by fine-tuning the initial value of $\varphi$ to be $\varphi_i\simeq 10^{-3}m_\varphi$, we can even obtain $\lambda=\mathcal{O}(1)$. We expect many effective field theories can satisfy the requirement of $\lambda$, including the Higgs boson in the SM of particle physics. We performed a numerical simulation, confirming that such an inflation process is a dynamic attractor that avoids the fine-tuning problem of standard inflation. 

We computed the primordial curvature perturbation spectrum of this KS inflation model. We found it is nearly scale-invariant and consistent with the current observation of E-modes in CMB, similar to the standard inflation, except the factor $c_s^{-1}\gg 1$. This large factor implies a more significant gap between the scalar and tensor modes of primordial metric perturbation for KS inflation. More specifically, by computing the tensor-to-scalar ratio, we found that compared to the standard inflation ($r_\textbf{standard}\simeq 0.04$), the current PGW searches ($r_\textbf{BICEP+}< 0.03$) favors the newly proposed KS inflation ($r_\textbf{KSI} \ll 0.04$). 

We want to emphasize that our estimation of the e-folding number of KS inflation is rough (we adopt a constant rolling velocity for $\varphi$ rather than the small quasi-oscillating rolling velocity to facilitate our estimation), which should be improved in the future. Furthermore, although the tensor-to-scalar ratio $r$ predicted by the standard inflation seems too large to be compatible with current PGW searches' results, there are still some mechanisms that can suppress a large primordial $r$ to be smaller during the (post-)reheating phase, for example, the perturbative resonance mechanism proposed in Ref.\cite{Li:2019std}. Therefore, a small $r$ implied by current observations may have multiple physics origins, which are worthy of further study. And the dynamic attractor is robust against various $V(\varphi)$ and $a(t)$. We expect it to apply multiple inflation and bouncing universe models in future studies (for an extensive review of cosmological models, see \cite{Nojiri:2017ncd} and citations to the original works), not only for KS inflation.


\acknowledgments
C.L. is supported by the NSFC under Grants No.11963005 and No.11603018, by Yunnan Provincial Foundation under Grants No.2016FD006 and No.2019FY003005, by Reserved Talents for Young and Middle-aged Academic and Technical Leaders in Yunnan Province Program, by Yunnan Provincial High-level Talent Training Support Plan Youth Top Program, and by the NSFC under Grant No.11847301 and by the Fundamental Research Funds for the Central Universities under Grant No. 2019CDJDWL0005. Y.C. is supported by the NSFC under Grants No.11775110.

\appendix

\section{The e-folding number}
\label{B}
Using Eq.(\ref{eq:ksiae}), we obtain the e-folding number of KS inflation,
\begin{equation}\label{eq:nedelt}
    N_e=\ln\left(\frac{a_r}{a_f}\right)=\frac{1}{2\sqrt{3}}\frac{m_\varphi^2}{\sqrt{\lambda}M_p}\Delta t~,
\end{equation}
where $\Delta t\equiv t_r-t_f$ being the duration of KS inflation. On the other hand, the change of $\varphi$ during KS inflation takes 
\begin{equation}
    \Delta \varphi= \varphi_v-\varphi_i\simeq \varphi_v=m_\varphi/\sqrt{2\lambda},
\end{equation}
where $\varphi_i$ being the value of $\varphi$ at the onset of KS inflation, and $\varphi_v$ being the expectation value of $\varphi$ at the true vacuum. 

According to our analysis and numerical result (Figure \ref{fig:phiattractor}), during KS inflation, the dynamic attractor locks $\varphi$ at a small velocity $\dot{\varphi}(t)\ll m_\phi^2$ rather than exponentially increasing. Before the tachyon fully condenses, $\varphi$ can freely roll along the potential and accelerates to $\dot{\varphi}_i$. After that, the fully condensed tachyon matter locks $\varphi$ around $\dot{\varphi}_i$, performing a quasi-oscillating rolling, as shown in Figure \ref{fig:phiattractor}. In this work, we approximate $\varphi(t)$ to be $\dot{\varphi}_i$ for simplicity. Thus we have      
\begin{equation}
    \Delta\varphi=\int_{t_i}^{t_r}\dot{\varphi}(t)dt\simeq\dot{\varphi}_i\Delta t= \dot{\varphi}_i\times 2\sqrt{3}\frac{\sqrt{\lambda}M_p}{m_\varphi^2}N_e=\frac{m_\varphi}{\sqrt{2\lambda}}~,
\end{equation}
which leads to
\begin{equation} \label{eq:neovdtvp}
    N_e=\frac{1}{2\sqrt{6}}\frac{1}{\lambda}\frac{m_\varphi}{M_p}\frac{m_\varphi^2}{\dot{\varphi}_i}~,
\end{equation}
where Eq.(\ref{eq:nedelt}) being used.

Substituting the locking condition $\ddot{\varphi}_i=0$ into the equation of motion of $\varphi$ during the annihilating D-brane and anti-D-brane pair dominated inflation phase (Phase II),
\begin{equation}
    \ddot{\varphi}+3H\dot{\varphi}+\frac{m_s^4}{V_0}\left(-2m_\varphi^2\varphi\right)=0~,
\end{equation}
we obtain
\begin{equation}\label{eq:dtvpvp}
    \dot{\varphi}_i\simeq\frac{2}{3}\frac{m_\varphi^2m_s^4}{HV_0}\varphi=\frac{2}{\sqrt{3}}\frac{m_\varphi^2}{\sqrt{m_s M_p}}\varphi_i~,
\end{equation}
which is the same order of Eq.(\ref{eq:vpdvpda}). Substituting Eq.(\ref{eq:dtvpvp}) into Eq.(\ref{eq:neovdtvp}), we obtain
\begin{equation}\label{eq:Neivvpi}
     N_e=\frac{1}{4\sqrt{2}}\frac{1}{\lambda}\sqrt{\frac{m_s}{M_p}}\left(\frac{m_\varphi}{\varphi_i}\right)~,
\end{equation}
which indicates that by fine-tuning the initial value of $\varphi$, one can obtain an arbitrarily large e-folding number for KS inflation ($N_e \propto \varphi_i^{-1}$). However, in generic, we expect the initial value of $\varphi$ is the same order of its mass,
\begin{equation}\label{eq:invpmvp}
    \varphi_i\simeq m_\varphi.
\end{equation}
Substituting Eq.(\ref{eq:invpmvp}) into Eq.(\ref{eq:Neivvpi}), we obtain 
\begin{equation}\label{eq:Nefinal}
    N_e=\frac{1}{4\sqrt{2}}\frac{1}{\lambda}\sqrt{\frac{m_s}{M_p}}~.
\end{equation}
Eq.(\ref{eq:Nefinal}) reflects that a small $\lambda$ corresponds to a large $\rho_\varphi$ and a large $\varphi_v$, suggests a large e-folding number for KS inflation.

\section{The equation of motion of primordial curvature perturbation during KS inflation}
\label{C}

In this section, we use the  perturbative Einstein equation 
\begin{equation}\label{eq:pereife}
\delta G_\nu^\mu=8\pi G\delta T_\nu^\mu
\end{equation}
to derive the equation of motion of primordial curvature perturbation (PCP) during the KS inflation,
\begin{equation}\label{eq:phikeqfullap}
\ddot{\Phi}_k+\left(H-\frac{\ddot{H}}{\dot{H}}\right)\dot{\Phi}_k+\left(2\dot{H}-H\frac{\ddot{H}}{\dot{H}}\right)\Phi_k+\frac{c_s^2k^2}{a^2}\Phi_k=0~,
\end{equation}
where $c_s^2=\left(1-\dot{\mathcal{T}}^2-\dot{\varphi}^2\right)$ being the sound speed of coupled tachyon matter, $\Phi_k$ being the Fourier modes of metric perturbation $\Phi(x,t)$, and $\Psi=-\Phi$ being used.

Using FLRW metric (Eq.(\ref{eq:flrwall})), we obtain the terms on the left-hand side (LHS) of Eq.(\ref{eq:pereife})
\begin{eqnarray}
&& \delta G^0_0=-2\frac{\nabla^2}{a^2}\Phi+6H(\dot{\Phi}+H\Phi),\label{eq:delgzz} \\
&& \delta G^0_i=-2\partial_i(\dot{\Phi}+H\Phi), \label{eq:delgzi}\\
&& \delta G^i_j=0, \label{eq:delgij}\\
&& \delta G^i_i=2\ddot{\Phi}+8H\dot{\Phi}+2(2\dot{H}+3H^2)\Phi~\label{eq:delgii}.
\end{eqnarray}

Expanding the right-hand side (RHS) term of Eq.(\ref{eq:efeforcth}) to the first order and using $\delta T_0^0=-\delta\rho$, $\delta T_i^0=-\partial_i\delta q$, $\delta T_i^j=\partial_i \partial^j\delta s$ and $\delta T_i^i=\delta p$, respectively, we obtain 
\begin{eqnarray}
&&\delta\rho
=\rho_\mathcal{T}\left(\frac{\dot{\varphi}\delta\dot{\varphi}+\dot{\mathcal{T}}\delta\dot{\mathcal{T}}-\left(\dot{\varphi}^2+\dot{\mathcal{T}}^2\right)\Phi}{1-\dot{\mathcal{T}}^2-\dot{\varphi}^2}+\frac{d\ln V(\mathcal{T})}{d\mathcal{T}}\delta \mathcal{T}\right)+\frac{dV(\varphi)}{d\varphi}\delta\varphi~, \label{eq:deltzz}\\
&& \delta  q=\rho_\mathcal{T}\left(\dot{\mathcal{T}}\delta\mathcal{T}+\dot{\varphi}\delta\varphi\right)~, \label{eq:deltzi}\\
&& \delta s=0~,\\ \label{eq:deltij}
&& \delta p
=p_\mathcal{T} \left(-\frac{\dot{\varphi}\delta\dot{\varphi}+\dot{\mathcal{T}}\delta\dot{\mathcal{T}}-\left(\dot{\varphi}^2+\dot{\mathcal{T}}^2\right)\Phi}{1-\dot{\mathcal{T}}^2-\dot{\varphi}^2}+\frac{d\ln V(\mathcal{T})}{d\mathcal{T}}\delta \mathcal{T}\right)-\frac{dV(\varphi)}{d\varphi}\delta\varphi \label{eq:deltii}
\end{eqnarray}
where $\rho_\mathcal{T}$ and $p_\mathcal{T}$ are the background energy density and pressure of the coupled tachyon matter, respectively.

Using  Eq.(\ref{eq:deltzi}), Eq.(\ref{eq:delgzz}), Eq.(\ref{eq:delgii}), Eq.(\ref{eq:deltzz}), and Eq.(\ref{eq:deltii}),  we have
\begin{align}
\frac{1}{8\pi G}\left(\delta G_i^i+c_s^2 \delta G_0^0\right)&=2\frac{d \ln V(\mathcal{T})}{\dot{\mathcal{T}} d\mathcal{T}}\left(-c_s^2\right)\delta q-2\frac{\ln V(\mathcal{T})}{d\mathcal{T}}p_\mathcal{T}\frac{\dot{\varphi}}{\dot{\mathcal{T}}}\delta \varphi-\frac{dV(\varphi)}{d\varphi}\delta\varphi~,\\
&=2\frac{d \ln V(\mathcal{T})}{\dot{\mathcal{T}} d\mathcal{T}}\left(-c_s^2\right)\delta q-\frac{\ln V(\mathcal{T})}{d\mathcal{T}}p_\mathcal{T}\frac{\dot{\varphi}}{\dot{\mathcal{T}}}\delta \varphi~~,\\
&=2\frac{d \ln V(\mathcal{T})}{\dot{\mathcal{T}} d\mathcal{T}}\left(-c_s^2\right)\delta q~. \label{eq:finaldq}
\end{align}
In the second line, we have used $-\frac{dV(\varphi)}{d\varphi}=\frac{d\ln V(\mathcal{T})}{d\mathcal{T}}p_\mathcal{T}\frac{\dot{\varphi}}{\dot{\mathcal{T}}}$, which is derived from the background part of Eq.(\ref{eq:eofmt}) and Eq.(\ref{eq:eofmvp}) during the KS inflation ($\ddot{T}\simeq0$ and $\ddot{\varphi}\simeq 0$). In the third line, we further neglect the term $\frac{\ln V(\mathcal{T})}{d\mathcal{T}}p_\mathcal{T}\frac{\dot{\varphi}}{\dot{\mathcal{T}}}\delta \varphi$ as it is suppressed by the factor $\dot{\varphi}/\dot{\mathcal{T}}\ll m_\varphi^2/m_s^2\simeq 0$. 

Substituting Eq.(\ref{eq:delgzz}), Eq.(\ref{eq:delgzi}), Eq.(\ref{eq:delgii}) and Eq.(\ref{eq:deltzi}) into Eq.(\ref{eq:finaldq}), we obtain
\begin{align}\label{eq:ddphialbeeq}
\ddot{\Phi}+\alpha\dot{\Phi}+\beta\Phi-c_s^2\frac{\nabla^2}{a^2}\Phi=0~,
\end{align}
where
\begin{align}
\alpha & \equiv H+(3H+3Hc_s^2)+2\frac{d \ln V(\mathcal{T})}{\dot{\mathcal{T}} d\mathcal{T}}c_s^2\simeq  H+3H ~;\\
\beta & \equiv 2\dot{H}+H\left(3H+3Hc_s^2+2\frac{d \ln V(\mathcal{T})}{\dot{\mathcal{T}} d\mathcal{T}}c_s^2\right)\simeq 2\dot{H}+H\cdot 3H~.
\end{align}

Submitting  
\begin{align}\label{eq:ddHdHcseq}
\frac{\ddot{H}}{\dot{H}}=\frac{\dot{\rho}+\dot{p}}{\rho+p}=\frac{\dot{\rho}_\mathcal{T}+\dot{p}_\mathcal{T}}{\rho_\mathcal{T}+p_\mathcal{T}}\simeq \frac{\dot{\rho}_\mathcal{T}}{\rho_\mathcal{T}}=-3H~,
\end{align} 
into Eq.(\ref{eq:ddphialbeeq}), we obtain
\begin{align}\label{eq:fullfinalphieq}
\ddot{\Phi}+\left(H-\frac{\ddot{H}}{\dot{H}}\right)\dot{\Phi}+\left(2\dot{H}-H\frac{\ddot{H}}{\dot{H}}\right)\Phi-c_s^2\frac{\nabla^2}{a^2}\Phi=0~,
\end{align}
which becomes Eq.(\ref{eq:phikeqfull}) after Fourier transformation. 

Note that in the standard single field slow-roll inflation model ($\mathcal{L}=-\frac{1}{2}\partial_\mu\phi\partial^\mu\phi+V(\phi)$), one has $c_s^2(\phi)=1$ and
\begin{align}
\alpha_\phi & = H+(3H+3Hc_s^2(\phi))+2\frac{d V(\mathcal{\phi})}{\dot{\phi} d\phi}c_s^2(\phi)=  H+6H+2\frac{d V(\mathcal{\phi})}{\dot{\phi} d\phi} ~;\\
\beta_\phi & = 2\dot{H}+H\left(3H+3Hc_s^2(\phi)+2\frac{d V(\mathcal{T})}{\dot{\phi} d\phi} c_s^2(\phi)\right)= 2\dot{H}+H\cdot \left(6H+2\frac{d V(\mathcal{\phi})}{\dot{\phi} d\phi} \right).
\end{align}
for Eq.(\ref{eq:ddphialbeeq}) \cite{Mukhanov:1990me}. Using  
\begin{equation}\label{eq:ddhodhstin}
\frac{\ddot{H}}{\dot{H}}=-6H-2\frac{d V(\mathcal{\phi})}{\dot{\phi} d\phi}, 
\end{equation}
derived in the standard inflation \cite{Mukhanov:1990me}, one can re-obtain 
\begin{align}\label{eq:fullfinalsfsl}
\ddot{\Phi}+\left(H-\frac{\ddot{H}}{\dot{H}}\right)\dot{\Phi}+\left(2\dot{H}-H\frac{\ddot{H}}{\dot{H}}\right)\Phi-\frac{\nabla^2}{a^2}\Phi=0~,
\end{align}
which leads to the desired nearly scale-invariant perturbation power spectrum. 

Comparing Eq.(\ref{eq:ddHdHcseq}) and Eq.(\ref{eq:ddhodhstin}), we find that, although $\ddot{H}/\dot{H}$ are different between the standard inflation model and the newly proposed KS model, Eq.(\ref{eq:fullfinalphieq}) and Eq.(\ref{eq:fullfinalsfsl}) are the same for them, expect $c_s^2\ll 1$ for KS inflation. This observation can explain why KS inflation also produces a nearly scale-invariant PCP from a mathematical viewpoint.


\end{document}